\documentclass[a4paper]{article}
\usepackage[left=20mm, top=20mm, right=20mm, bottom=20mm, nohead]{geometry}
\usepackage[T1,T2A]{fontenc}
\usepackage[utf8]{inputenc}
\usepackage[english]{babel}
\usepackage{graphicx}
\usepackage{hyperref}
\usepackage{cancel}
\usepackage{amsmath}
\usepackage{amssymb}
\usepackage{enumitem}
\usepackage{bm}
\usepackage{verbatim}
\usepackage{multirow}
\usepackage[labelsep=period]{caption}

\newcommand{\pder}[2]{\frac{\partial #1}{\partial #2}}
\newcommand{\dpder}[2]{\frac{\partial^2 #1}{\partial {#2}^2}}
\newcommand{\grad}[1]{{\, \text {grad} #1}}

\let\kappa=\varkappa
\let\epsilon=\varepsilon
\let\phi=\varphi
\renewcommand{\div}[1]{\, \text {div} #1}
\newcommand{\rot}[1]{\, \text {rot} #1}
\newcommand{\norm}[1]{\left\lVert#1\right\rVert}
\newcommand{\real}{\operatorname{Re}}
\graphicspath{{imgs/}}

\title{
\vspace{-2cm} 
\begin{flushright}
{\normalsize INR-TH-2021-018}
\end{flushright}
\vspace{0.5cm}
On resonant generation of electromagnetic modes
\\ in nonlinear electrodynamics: \\
Classical approach}

\author{Ilia Kopchinskii$^{1,2}$\thanks{{\bf e-mail}: 	kopchinskii@ms2.inr.ac.ru}$\;$, Petr Satunin$^2$ \\
\normalsize\it $^1$ Moscow State University, \\ 
\normalsize\it Leninskiye Gory, 119991 Moscow, Russia \\
\normalsize\it $^2$ Institute for Nuclear Research of the Russian Academy
of Sciences, \\  
\normalsize\it 60th October Anniversary Prospect, 7a, 117312  Moscow, Russia}
\date{}

\begin{document}
\thispagestyle{empty}
\maketitle

\begin{abstract}
The paper explores a theoretical possibility of resonant amplification of electromagnetic modes generated by a nonlinear effect in Euler-Heisenberg electrodynamics. Precisely, we examine the possibility of the amplification for the third harmonics induced by a single electromagnetic mode in radiofrequency cavity, as well as the generation of signal mode of combined frequencies induced by two pump modes ($\omega_1$ and $\omega_2$) in the cavity. Solving inhomogeneous wave equations for the signal mode, we formulate two resonant conditions for a cavity of arbitrary shape, and apply the obtained formalism to linear and rectangular cavities. We explicitly show that the third harmonics as well as the mode of combined frequency $2\omega_1 + \omega_2$ are not resonantly amplified while the signal mode with frequency $2\omega_1 - \omega_2$ is amplified for a certain cavity geometry.
\end{abstract}

\section{Introduction}

The self-interaction of an electromagnetic field, being absent in classical theory, appears in quantum theory due to radiative corrections which include the contribution of virtual electrons. At low frequency of the electromagnetic field the quantum effect is described in terms of the effective Euler-Heisenberg Lagrangian \cite{Heisenberg:1935qt,Euler:1935zz} (the detailed historical review is presented in \cite{Dunne:2012vv}). The most distinctive effect of Euler-Heisenberg nonlinear electrodynamics is the process of photon-photon scattering \cite{Heisenberg:1935qt}. Other nonlinear electrodynamics effects include vacuum birefringence and dichroism for an electromagnetic wave in  classical intense electromagnetic background \cite{Baier:1967zza,Bialynicka-Birula:1970nlh}. Besides the Euler-Heisenberg contribution, effective nonlinear interactions in electrodynamics appear if the full theory contains scalar or pseudoscalar particles interacting with electromagnetic field \cite{Evans:2018qwy}.

No effect, predicted in 
nonlinear electrodynamics, has been experimentally observed to this moment. The reason is the extreme smallness of the  self-coupling for the electromagnetic field. Nevertheless, several experimental attempts to probe it with high-intensity electromagnetic fields have been made. The experiment which comes closest to the Euler-Heisenberg limit is the polarization experiment with intensive laser fields PVLAS \cite{DellaValle:2014xoa,DellaValle:2015xxa}. The final PVLAS experimental sensitivity to photon self-coupling is one order of magnitude weaker than the prediction of Euler-Heisenberg \cite{DellaValle:2015xxa}.

Another experimental proposal referred to high-intensity electromagnetic modes in cavities instead of laser fields. The idea of such experiment was suggested in the early 2000s \cite{Brodin:2001zz,Eriksson:2004cz}. The proposed setup consists of a single superconducting cavity filled with two non-equal ``pump'' modes. In the presence of self-interaction, one expects an excitation of a third ``signal'' mode whose frequency is a linear combination of the pump modes' frequencies. Due to the smallness of nonlinear effect the signal mode can be detected only if it is resonantly amplified. The application of a single cavity setup to the searching for pseudoscalar axion-like particles was proposed in \cite{Bogorad:2019pbu}\footnote{The generalization to scalar particles and CP-violating term was considered in \cite{Gorghetto:2021luj}.}: if the particle is heavy (the mass is much greater than the frequencies of the pump modes)\footnote{In the case of small mass of an axion-like particle the similar experiments with two cavities have been proposed \cite{Janish:2019dpr, Gao:2020anb, Salnikov:2020urr}.}, the problem is reduced to the aforementioned nonlinear electrodynamics.

In papers \cite{Brodin:2001zz,Eriksson:2004cz, Bogorad:2019pbu} the solutions of nonlinear wave equations describing the resonant growth of a signal mode have not been provided explicitly. In a recent paper \cite{Shibata:2020don} that nonlinear wave equation was exactly solved in a simplified one-dimensional model. It was shown that, contrary to the naive estimates, signal mode with triple frequency is not resonantly generated in a one-dimensional ``cavity''; the only resonant amplification is observed at the pump mode's frequency. The goal of current article is to generalize these calculations to realistic three-dimensional cavities.

The paper is organized as follows. Section \ref{sec:nonlinear-maxwell} is devoted to nonlinear Maxwell and wave equations. In Section \ref{sec:formalism} we introduce our general formalism of searching for resonant modes in arbitrary cavity. In Section \ref{sec:1d-cavity} we apply the formalism to one-dimensional cavity filled by one or two pump modes. In Section \ref{sec:3d-cavity} we generalize the results of the previous section to three-dimensional rectangular cavity. In Section \ref{sec:discuss} we discuss obtained results.

\section{Nonlinear Maxwell and wave equations}
\label{sec:nonlinear-maxwell}

In this section we briefly review the field equations appeared in nonlinear electrodynamics. The Euler-Heisenberg Lagrangian in the limit of weak electric and magnetic field ($E, H \ll m_e^2/e\,\sim\, 10^{18}\,\mbox{V/m}$) takes the form \cite{Heisenberg:1935qt,Euler:1935zz}:
\begin{equation}
\label{eq:lagrangian}
\mathcal{L} = -\frac{1}{4} \mathcal{F} + \epsilon (\mathcal{F}^2 + \beta \mathcal{G}^2), \qquad \epsilon = \frac{\alpha_e^2}{90 m_e^4}, \quad \beta = \frac{7}{4},
\end{equation}
where $\alpha_e$ is fine structure constant and $m_e$ is the electron mass. In presence of hypothetical scalar or pseudoscalar particles in theory the coefficients $\epsilon$ and $\beta$ are modified \cite{Evans:2018qwy, Gorghetto:2021luj}. The electromagnetic field invariants have the standard form, 
\begin{equation}
\label{eq:invariants}
\mathcal{F} \equiv F_{\mu\nu} F^{\mu\nu} = -2 \left(E^2 - H^2\right),\qquad \qquad \mathcal{G} \equiv  F_{\mu\nu} {\widetilde F}^{\mu\nu} = -4 \left(\bm E \cdot \bm H\right).
\end{equation}

The electromagnetic field equations obtained from the Lagrangian (\ref{eq:lagrangian}) have the form analogous to Maxwell equations in medium \cite{Euler:1935zz},
\begin{equation}
\label{eq:mod-maxwell}
\begin{aligned}
\rot{\bm H} &= \pder{\bm E}{t} + \left[\pder{\bm P}{t} - \rot{\bm M}\right], \\
\rot{\bm E} &= -\pder{\bm H}{t}, \\
\end{aligned}
\hspace{2cm}
\begin{aligned}
\div{\bm H} &= 0, \\
\div{\bm E} &= \left[-\div{\bm P}\right],
\end{aligned}
\end{equation}
where  ${\bm P}$ and ${\bm M}$ denote vacuum polarization and magnetization respectively,
\begin{equation}
\label{eq:corrections}
\begin{aligned}
\bm P(\bm x, t) &\equiv 16 \epsilon \left[\left(E^2 - H^2\right) \bm E + 2 \beta (\bm E \cdot \bm H) \bm H\right], \\
\bm M(\bm x, t) &\equiv 16 \epsilon \left[\left(E^2 - H^2\right) \bm H - 2 \beta (\bm E \cdot \bm H) \bm E\right].
\end{aligned}
\end{equation}

The field equations \eqref{eq:mod-maxwell} yield to modified wave equations both for amplitudes for electric and magnetic fields,
\begin{equation}
\label{eq:mod-waves}
\begin{aligned}
\Box \bm E &= \pder{}{t} \rot{\bm M} + \grad{\div{\bm P}} - \dpder{\bm P}{t}, \\
\Box \bm H &= \pder{}{t} \rot{\bm P} - \grad{\div{\bm M}} + \Delta \bm M.
\end{aligned}
\end{equation}
Note that the plane electromagnetic wave is a solution of modified wave equations \eqref{eq:mod-waves} since both electromagnetic invariants vanish at the plane wave configuration, $\mathcal{F}=\mathcal{G}=0$. However, a linear combination of plane waves is no longer a solution of equations \eqref{eq:mod-waves} and so becomes unstable, leading to the production of new modes.

\section{General formalism of searching for resonant modes}
\label{sec:formalism}

One of the interesting features of modified wave equations \eqref{eq:mod-waves} is the possibility for generation of higher-order harmonics by one or two initial electromagnetic modes in vacuum. Traditionally \cite{Brodin:2001zz,Eriksson:2004cz,Bogorad:2019pbu}, we consider the following setup devoted to the search for higher-order harmonics. We take a superconducting radiofrequency cavity (SRF) filled with one or two ``pump'' modes of very high amplitudes $\bm E^{pump}, \bm H^{pump}$, and look for generation of a ``signal'' mode of different frequency with amplitudes $\bm E^{sig}, \bm H^{sig}$ which are expected to be small due the smallness of nonlinear coupling coefficient $\epsilon$. Treating the signal mode as a small perturbation in eq.~\eqref{eq:mod-waves} and assuming the hierarchy of scales $|\bm E^{sig}| \sim \epsilon \left( |\bm E^{pump}|\right)^3 \ll |\bm E^{pump}|$, one obtains in the zeroth order trivial wave equations for the pump modes $\Box \bm E^{pump} = 0, \ \Box \bm H^{pump} = 0$, and in the first order:
\begin{gather}
\begin{aligned}
\Box \bm E^{sig} &=  \pder{}{t} \rot{\bm M(\bm E^{pump}, \bm H^{pump})} + \grad{\div{\bm P(\bm E^{pump}, \bm H^{pump})}} - \dpder{\bm P(\bm E^{pump}, \bm H^{pump})}{t}, \\
\Box \bm H^{sig} &=  \pder{}{t} \rot{\bm P(\bm E^{pump}, \bm H^{pump})} - \grad{\div{\bm M(\bm E^{pump}, \bm H^{pump})}} + \Delta \bm M(\bm E^{pump}, \bm H^{pump}).
\end{aligned}
\label{eq:linear}
\end{gather}
Here the polarization and magnetization vectors \eqref{eq:corrections} are computed on the pump mode configuration. Instead of nonlinear eq.~\eqref{eq:mod-waves}, eq.~\eqref{eq:linear} is a linear wave equation on the signal mode amplitudes with nonzero r.h.s.~The solution of eq.~\eqref{eq:linear} determines the evolution of the signal mode at classical level.

The equations \eqref{eq:linear} are to be solved in a given cavity $D$ of finite volume. Furthermore, in order to take into account small dissipation, we introduce the dissipative term which includes the first-order time derivative and the damping coefficient $\Gamma$:
\begin{equation}
\left\{
\begin{aligned}
\left(\Box - \Gamma\partial_t\right) \bm E^{sig}(\bm x,t) &= \bm F(\bm x, t), \quad \bm x \in D, ~t > 0, \\
\bm E^{sig}(\bm x, 0) &= 0, \quad \bm x \in D, \\
\bm n \times \bm E^{sig}(\bm x,t)& = 0, \quad  {\bm x \in S}.
\end{aligned}
\right. \label{eq:problem}
\end{equation}
Here $S$ denotes the surface of the cavity $D$, $\bm n$ is the normal to the surface $S$. $\bm F(\bm x, t)$ denotes the r.h.s.~of the electric equation in \eqref{eq:linear}. The boundary conditions refer to an ideal conducting surface. The similar system should be written for magnetic component of the signal mode $\bm H^{sig}$, however we will further skip it for the sake of shortness.

The signal field $\bm E^{sig}(\bm x, t)$ can be expanded into the cavity eigenmodes,
\begin{equation}
    \bm E^{sig} (\bm x,t) = \sum_k  E^{sig}_k (t) \, \bm{\mathcal E}_k (\bm x).
    \label{eq:expansion}
\end{equation}
Here $\bm{\mathcal E}_k (\bm x)$ are the full system of eigenfunctions with eigenvalues $\omega_k$, satisfying the equation $\left(\Delta + \omega_k^2\right) \bm{\mathcal E}_k (\bm x) = 0$ and boundary conditions given in the last line of eq.~\eqref{eq:problem}. Substituting expansion \eqref{eq:expansion} to the first eq.~of \eqref{eq:problem} and integrating over the whole cavity with a mode $\bm {\mathcal E}_n(\bm x)$, one obtains:
\begin{equation}
   \ddot{E}^{sig}_n(t) + \Gamma\dot{E}^{sig}_n(t) + \omega_n^2E^{sig}_n(t) = F_n(t), \qquad F_n(t) \equiv \frac{\int_D dV \bm F(\bm x,t) \cdot \bm {\mathcal E}_n(\bm x)}{\int_D dV \bm {\mathcal E}_n^2(\bm x)} \equiv \frac{(\bm F, \bm {\mathcal E}_{n})}{\norm{\bm {\mathcal E}_{n}}^2}. 
   \label{eq:ode}
\end{equation}
Here we made the following notations: $\int_D dV$ for the integration over the volume of the cavity $D$,  $(\bm F, \bm {\mathcal E})$ for the inner product of two vector functions $\bm F$ and $\bm {\mathcal E}$, and $\norm{\mathcal{E}}=\sqrt{(\mathcal{E},\mathcal{E})}$ for the norm of function $\mathcal{E}$. All frequency components of $F_n(t)$ lead to generation of a signal mode with a certain amplitude. If $F_n(t)$ includes the component  $F_n(t) \supseteq \real\left(F_n^0 e^{-i\omega t}\right)$, the signal field of the amplitude 
\begin{equation}
    E^{sig}_n(t) = \real \frac{F_n^0 e^{-i\omega t}}{-\omega^2 -i\omega \Gamma + \omega_n^2}
    \label{sol}
\end{equation}
is generated in the steady regime. If the frequency $\omega$ coincides with one of the cavity eigenfrequencies, $\omega = \omega_n$, the first and the third terms in the denominator of \eqref{sol} cancel each other and so the amplitude of such signal mode is resonantly enhanced, $E^{sig}_n(t) = \real\left(i  F_n^0 e^{-i\omega t}/(\Gamma \omega) \right)$.

Let us summarize aforementioned expressions in a more strict way as a criterion for resonant amplification of the signal mode. Assume that the r.h.s.~of one of the six scalar wave equations \eqref{eq:linear} contains the frequency component $\omega_{sig}$. The signal mode with the frequency $\omega_{sig}$ is resonantly amplified if both of the following conditions hold simultaneously,
\begin{enumerate}
\item{Frequency $\omega_{sig}$ belongs to the cavity spectrum ($\exists ~ n \in \mathbb{N}: ~ \omega_{sig} = \omega_n $),}
\item{The scalar product $F_n(t)$ of the r.h.s.~of the considered scalar wave equation from (\ref{eq:linear}) with the $n$-th cavity eigenmode contains the frequency component $\omega_{sig}=\omega_n$.}
\end{enumerate}
Note that even if  $\bm F(\bm x, t)$ contains a frequency component $\omega_m$, it may disappear from $F_m(t)$ due to the integration with orthogonal cavity mode. An example supporting this statement will be provided in the next section.

\section{One-dimensional cavity}
\label{sec:1d-cavity}
In this section we consider a model of one-dimensional cavity directed along the $Ox$ axis,  $D = (0, a)$. The $y$ and $z$ dimensions of the cavity are assumed to be significantly larger than the x-dimension, $L_y, L_z \gg L_x \equiv a$. The cavity system\footnote{For mathematical completeness one has to add a constant mode (with zero frequency) to the system \eqref{eq:eigen-lin}.} of eigenfunctions assuming ideal conducting walls takes a simple form:
\begin{equation}
\left\{\begin{aligned}
\bm{\mathcal{E}}_n(x) &= \sin(k_n x) e^{-i\frac{\pi}{2}} \, \bm e_y \\
\bm{\mathcal{M}}_n(x) &= \cos(k_n x) \, \bm e_z
\end{aligned}\right., \qquad k_n = \frac{\pi n}{a}, \quad \norm{\bm{\mathcal{E}}_n}^2 = \norm{\bm{\mathcal{M}}_n}^2 = \frac{a}{2}, \qquad n \in \mathbb{N}.
\label{eq:eigen-lin}
\end{equation} 
The dynamics of a cavity mode with wavenumber $k_n$ is just an oscillation with frequency $\omega_n=k_n$,
\begin{equation}
\left\{
\begin{aligned}
\bm E^{pump}(x, t) &= F_0 \real\left[\bm{\mathcal{E}}_{n}(x) e^{i \omega_n t}\right] = F_0 \sin(\omega_n x) \sin(\omega_n t) \, \bm e_y, \\
\bm H^{pump}(x, t) &= F_0 \real\left[\bm{\mathcal{M}}_{n}(x) e^{i \omega_n t}\right] = F_0 \cos(\omega_n y) \cos(\omega_n t) \, \bm e_z.
\end{aligned}
\right. \label{eq:pumps-lin-1}
\end{equation}

\subsection{Single pump mode}
First we consider an excitation of the one-dimensional cavity with a single pump mode of frequency $\omega_n$, see (\ref{eq:pumps-lin-1}).
At the pump mode configuration (\ref{eq:pumps-lin-1}) the invariant  $\mathcal{F} \neq 0$ while the second invariant $\mathcal{G}$ vanishes. Substituting the pump mode fields \eqref{eq:pumps-lin-1} to the expression for the inhomogeneities of nonlinear wave equation \eqref{eq:linear} and performing a simple but cumbersome trigonometric calculation\footnote{The calculations were additionally verified in the computer algebra system \textsc{wxMAXIMA 21.02.0}, see \cite{Maxima}.}, we obtain inhomogeneous wave equations for signal modes (cl. \cite{Shibata:2020don}),
\begin{equation}
\small
\begin{aligned}
\left(\Box - \Gamma\partial_t\right) \bm E^{sig} &= 8\epsilon F_0^3\omega_n^2 \Bigl[2\sin(\omega_n x)\sin(\omega_n t) + \sin(3\omega_n x)\sin(\omega_n t) - 3 \sin(\omega_n x)\sin(3\omega_n t)\Bigr] \, \bm e_y= \bm F^{el}(x,t), \\
\left(\Box - \Gamma\partial_t\right) \bm H^{sig} &= 8\epsilon F_0^3\omega_n^2 \Bigl[2\cos(\omega_n x)\cos(\omega_n t) + 3\cos(3\omega_n x)\cos(\omega_n t) - \cos(\omega_n x)\cos(3\omega_n t)\Bigr] \, \bm e_z = \bm F^{mg}(x,t).
\end{aligned}
\label{eq:rhs-lin-1}
\end{equation}
Note that both equations \eqref{eq:rhs-lin-1} contain terms $\sin(\omega_n x)\sin(\omega_n t)$ or $\cos(\omega_n x)\cos(\omega_n t)$, which obviously result in a resonant enhancement of the signal mode with frequency $\omega_n$. However, the r.h.s.~of \eqref{eq:rhs-lin-1} do not contain terms like $\sin(3\omega_n x)\sin(3\omega_n t)$, which would produce a signal mode of triple frequency. Formally, let us use the resonance criterion formulated in the previous section. The projections of the r.h.s.~of \eqref{eq:rhs-lin-1} on the cavity eigenfunctions are:
\begin{equation*}
\begin{aligned}
F^{el}_{n}(t) &\equiv \frac{(\bm F^{el}, \bm {\mathcal E}_{n})}{\norm{\bm {\mathcal E}_{n}}^2} = \frac{2}{a}\int\limits_0^a F^{el}(x,t) \sin(\omega_n x) dx = 8\epsilon F_0^3\omega_n^2 \Bigl[2\sin(\omega_n t) - 3\sin(3\omega_n t)\Bigr], \\
F^{mg}_{n}(t) &\equiv \frac{(\bm F^{mg}, \bm{\mathcal M}_{n})}{\norm{\bm{\mathcal M}_{n}}^2} = \frac{2}{a}\int\limits_0^a F^{mg}(x,t) \cos(\omega_n x) dx = 8\epsilon F_0^3\omega_n^2 \Bigl[2\cos(\omega_n t) - \cos(3\omega_n t)\Bigr], \\
F^{el}_{3n}(t) &\equiv \frac{(\bm F^{el}, \bm{\mathcal E}_{3n})}{\norm{\bm{\mathcal E}_{3n}}^2} = \frac{2}{a}\int\limits_0^a F^{el}(x,t) \sin(3\omega_n x) dx = 8\epsilon F_0^3\omega_n^2 \Bigl[\sin(\omega_n t)\Bigr], \\
F^{mg}_{3n}(t) &\equiv \frac{(\bm F^{mg}, \bm{\mathcal M}_{3n})}{\norm{\bm{\mathcal M}_{3n}}^2} = \frac{2}{a}\int\limits_0^a F^{mg}(x,t) \cos(3\omega_n x) dx = 8\epsilon F_0^3\omega_n^2\Bigl[3\cos(\omega_n t)\Bigr].
\end{aligned}
\end{equation*}

As the amplitudes are of the same order (as small as $\epsilon F_0^3 \omega_n^2$), for the sake of shortness we organize the computed projections into a table.
\begin{table}[ht!]
\centering
\def\arraystretch{1.3}
\begin{tabular}{|l|c|c|}
    \hline
     & $n$ & $3n$ \\
    \hline
    $\bm F^{el}$ & $\omega_{n}, ~ \omega_{3n}$ & $\omega_{n}$ \\
    \hline
    $\bm F^{mg}$ & $\omega_{n}, ~ \omega_{3n}$ & $\omega_{n}$ \\
    \hline
\end{tabular}
\caption{Examination of the resonance criterion for a single pump mode in 1D-cavity.}
\label{tbl:single-1D}
\end{table}

The Table \ref{tbl:single-1D} is to be interpreted as follows: the upper row contains mode numbers (on which projections were computed), the leftmost column enumerates the r.h.s.~of non-homogeneous wave equations for signal modes. Then, every cell contains those frequencies, which have been found in a projection of the r.h.s.~from corresponding row onto an eigenmode with the number from corresponding column. In case of one-dimensional cavity being excited with a single pump mode, one can easily correlate the four projections computed right above with the four cells in the table. The understanding of how the examination of the resonance criterion is presented in the table will be important later for more complex configurations, where the direct calculations result in formulas too long to be listed entirely.

As we see from Table \ref{tbl:single-1D}, the new triple frequency does not belong to a spectrum of any projection onto the $3n$-eigenmode. Thus, the condition 2 of the resonance criterion is not satisfied and therefore the signal mode with triple frequency is not resonantly enhanced.

\subsection{Two pump modes}

The next configuration we consider is the excitation of one-dimensional cavity with two pump modes of frequencies $\omega_n$ and $\omega_p$. Since the linear cavity exposes rotational symmetry along axis $Ox$, we introduce an arbitrary angle $\alpha$ between the polarization planes of the pump modes:
\begin{equation}
\left\{
\begin{aligned}
\bm E^{pump}(x,t) &= F_0 \real\left[\bm{\mathcal{E}}_{n}(x) e^{i \omega_n t} + \bm{\hat R}_x(\alpha)\bm{\mathcal{E}}_{p}(x) e^{i \omega_p t}\right], \\
\bm H^{pump}(x,t) &= F_0 \real\left[\bm{\mathcal{M}}_{n}(x) e^{i \omega_n t} + \bm{\hat R}_x(\alpha)\bm{\mathcal{M}}_{p}(x) e^{i \omega_p t}\right],
\end{aligned}
\right. \qquad \bm{\hat R}_x(\alpha) =
\begin{pmatrix}
    1 & 0 & 0 \\
    0 & \cos\alpha & -\sin\alpha \\
    0 & \sin\alpha & \cos\alpha
\end{pmatrix}.
\label{eq:pumps-lin-2}
\end{equation}

The eigenmodes $\bm{\mathcal{E}}_{n}(x)$, $\bm{\mathcal{M}}_{n}(x)$ are given by \eqref{eq:eigen-lin}. In contrast to the case of a single pump mode, both electromagnetic invariants \eqref{eq:invariants} are nonzero at the current configuration. The inhomogeneous wave equations for the signal mode read,
\begin{equation}
\left(\Box - \Gamma\partial_t\right) \bm E^{sig}(x,t) = \bm F^{el}(x,t) = \left(\begin{array}{c} 0 \\ F^{el}_y \\ F^{el}_z \end{array}\right),
\qquad
\left(\Box - \Gamma\partial_t\right) \bm H^{sig}(x,t) = \bm F^{mg}(x,t) = \left(\begin{array}{c} 0 \\ F^{mg}_y \\ F^{mg}_z \end{array}\right),
\label{eq:rhs-lin-2}
\end{equation}
Here the inhomogeneities $\bm F^{el}(x,t)$ and $\bm F^{mg}(x,t)$ are calculated by the substitution of the field configuration \eqref{eq:pumps-lin-2}  to the general expression \eqref{eq:linear}. This calculation was performed in the computer algebra system \textsc{wxMAXIMA}; the resulting expressions are too long to be listed here explicitly, see \cite{Maxima}.

Nevertheless, the restrictions on the set of signal modes can be obtained taking a look at the structure of the inhomogeneities $\bm F^{el}(x,t)$ and $\bm F^{mg}(x,t)$. 
Since the expressions are cubic (see eq.~\eqref{eq:corrections}) relatively to the pump modes which are simple trigonometric functions, one can expect at most the following frequencies for signal modes in \eqref{eq:rhs-lin-2}:
\begin{equation}
    \omega_{sig} \in \{ \omega_n, ~\omega_p, ~3\omega_n, ~3\omega_p, ~2\omega_n \pm \omega_p, ~2\omega_p \pm \omega_n \}.
    \label{eq:possibles}
\end{equation}
The possible wavenumbers for the signal mode belong to the similar set,
\begin{equation}
    k_{sig} \in \{ k_n, ~k_p, ~3k_n, ~3k_p, ~2k_n \pm k_p, ~2k_p \pm k_n \}.
    \label{eq:possibleswn}
\end{equation}
It follows from the condition 1 of the criterion that the wavenumbers \eqref{eq:possibleswn} have to match with the corresponding frequencies, \eqref{eq:possibles}.

At the following step we examine the condition 2: we project the inhomogeneities of eqs.~\eqref{eq:rhs-lin-2} onto cavity eigenmodes, whose frequencies may hypothetically appear due to cubic nonlinearities \eqref{eq:possibles}. This stage of calculations was also carried out in \textsc{wxMAXIMA} system (see \cite{Maxima});  the results are presented in Table \ref{Table2} which is constructed analogously to that in the previous section.
\begin{table}[ht!]
\centering
\def\arraystretch{1.3}
\begin{tabular}{|l|c|c|c|c|c|c|c|c|}
    \hline
     & $n$ & $3n$ & $2n-p$ & $2n+p$ & $p$ & $3p$ & $2p-n$ & $2p+n$ \\
    \hline
    $\bm F^{el}$ & \multirow{2}{*}{$\omega_{n}, ~ \omega_{2p\pm n}, ~ \omega_{3n}$} & \multirow{2}{*}{$\omega_{n}$} & \multirow{2}{*}{$\omega_{p}, ~ \omega_{2n+p}$} & \multirow{2}{*}{$\omega_{p}, ~ \omega_{2n-p}$} & \multirow{2}{*}{$\omega_{p}, ~ \omega_{2n\pm p}, ~ \omega_{3p}$} & \multirow{2}{*}{$\omega_{p}$} & \multirow{2}{*}{$\omega_{n}, ~ \omega_{2p+n}$} & \multirow{2}{*}{$\omega_{n}, ~ \omega_{2p-n}$} \\
    \cline{1-1}
    $\bm F^{mg}$ & & & & & & & & \\
    \hline
\end{tabular}
\caption{Examination of the resonance criterion for two pump modes in 1D-cavity.}
\label{Table2}
\end{table}

The condition 2 of the resonance criterion is satisfied only for the signal frequencies $\omega_n$ and $\omega_p$, which are shadowed by the pump modes. Thus, in the case of two pump modes in one-dimensional cavity, resonant amplification of signal modes with mixed frequencies does not occur.

\section{Rectangular cavity}
\label{sec:3d-cavity}

In this section we proceed with a rectangular cavity $D = (0,a)\times(0,b)\times(0,c)$. Within the approximation of perfectly conducting walls the system of eigenfunctions separates into two subsets (relative to $Oz$ axis) --- TE-modes and TM-modes \cite[pp. 25--28]{Hill:2014}:
\begin{equation}
\begin{gathered}
\bm{\mathcal{E}}^{TM}_{npq}(\bm x), ~~
\bm{\mathcal{M}}^{TM}_{npq}(\bm x) \perp \bm e_z, \quad n,p \in \mathbb{N}, ~ q \in \mathbb{N}_0 \\
\bm{\mathcal{E}}^{TE}_{npq}(\bm x) \perp \bm e_z, ~~
\bm{\mathcal{M}}^{TE}_{npq}(\bm x), \quad n,p \in \mathbb{N}_0, ~ q \in \mathbb{N} \\
\end{gathered},
\qquad \bm k_{npq} = \left(\frac{\pi n}{a},\, \frac{\pi p}{b}, \, \frac{\pi q}{c}\right), \quad \norm{\mathcal{E}^i}^2 = \norm{\mathcal{M}^i}^2 = \frac{abc}{8}.
\label{eq:eigen-rect}
\end{equation}
Time evolution of $npq$-modes is an oscillation with the frequency $\omega_{npq} = |\bm k_{npq}| =  \pi \sqrt{\frac{n^2}{a^2} + \frac{p^2}{b^2} + \frac{q^2}{c^2}}$.

\subsection{Single pump mode}

Let us consider a single pump mode with an eigenfrequency $\omega_{npq}$. Since the division into TE- and TM-modes is purely artificial in case of rectangular cavity, we arbitrarily choose the TM$npq$ pump mode:
\begin{equation}
\left\{
\begin{aligned}
\bm E^{pump}(\bm x, t) &= F_0 \real\left[\bm{\mathcal{E}}^{TM}_{npq}(\bm x) e^{i \omega_{npq} t}\right], \\
\bm H^{pump}(\bm x, t) &= F_0 \real\left[\bm{\mathcal{M}}^{TM}_{npq}(\bm x) e^{i \omega_{npq} t}\right].
\end{aligned}
\right. \label{eq:pumps-rect-1}
\end{equation}

At the single mode configuration \eqref{eq:pumps-rect-1} the electromagnetic invariant $\mathcal{G}$ vanishes, similarly  to the one-dimensional case; the invariant $\mathcal{F}$ is still nonzero.
As previously, the r.h.s.~parts of the linearized wave equations are calculated using eqs.~\eqref{eq:corrections} and \eqref{eq:mod-waves}:
\begin{equation}
\left(\Box + \Gamma\partial_t\right) \bm E(\bm x,t) = \bm F^{el}(\bm x,t), \qquad
\left(\Box + \Gamma\partial_t\right) \bm H(\bm x,t) = \bm F^{mg}(\bm x,t).
\label{eq:rhs-rect-1}
\end{equation}

For arbitrary integers $(n,p,q)$ all components of the obtained inhomogeneities are non-zero. In order to examine the resonance criterion, their spatial projections on corresponding eigenfunctions are calculated (for instance, $\bm F^{el}$ is to be projected on $\bm {\mathcal E}^{TM}$ and $\bm {\mathcal E}^{TE}$). The temporal spectra of all non-vanishing projections are listed in the Table \ref{tab:3D-1-pump}.
\begin{table}[ht!]
\centering
\def\arraystretch{1.3}
\begin{tabular}{|l|c|c|c|c|c|c|c|c|}
    \hline
    TM$(\cdots)$ & $n,p,q$ & $3n,p,q$ & $n,3p,q$ & $n,p,3q$ & $n,3p,3q$ & $3n,p,3q$ & $3n,3p,q$ & $3n,3p,3q$ \\
    \hline
    $\bm F^{el}$ & \multicolumn{7}{|c|}{\multirow{2}{*}{$\omega_{npq}, ~ 3\omega_{npq}$}} & \multirow{2}{*}{$\omega_{npq}$} \\
    \cline{1-1}
    $\bm F^{mg}$ & \multicolumn{7}{|c|}{} & \\
    \hline \hline \hline
    TE$(\cdots)$ & $n,p,q$ & $3n,p,q$ & $n,3p,q$ & $n,p,3q$ & $n,3p,3q$ & $3n,p,3q$ & $3n,3p,q$ & $3n,3p,3q$ \\
    \hline
    $\bm F^{el}$ & \multicolumn{6}{|c|}{\multirow{2}{*}{$\omega_{npq}, ~ 3\omega_{npq}$}} & \multicolumn{2}{|c|}{\multirow{2}{*}{$\omega_{npq}$}} \\
    \cline{1-1}
    $\bm F^{mg}$ & \multicolumn{6}{|c|}{} & \multicolumn{2}{|c|}{} \\
    \hline
\end{tabular}
\caption{Examination of the resonance criterion for a single pump mode in rectangular cavity.}
\label{tab:3D-1-pump}
\end{table}

From the Table \ref{tab:3D-1-pump} one concludes that only the lowest frequency $\omega_{npq}$ is amplified, whilst the higher-order harmonics remain suppressed.
Note that, in contrast to the case of one-dimensional cavity, the modes with intermediate sets of wavenumbers (e.g.~$(n,3p,q)$) do appear in the rectangular cavity. However, neither the pump mode frequency nor the triple frequency fits these sets of wavenumbers (condition 1 of resonance criterion fails), so these modes are not resonantly amplified. 

\subsection{Two pump modes}

The last configuration being considered includes two pump modes (for certainty, one TM- and one TE-mode) excited in a rectangular cavity. The electric and magnetic fields of this configuration read,
\begin{equation}
\begin{aligned}
\bm E^{pump}(\bm x, t) &= F_0 \real\left[\bm{\mathcal{E}}^{TM}_{n_1,p_1,q_1}(\bm x) e^{i \omega_1 t} + \bm{\mathcal{E}}^{TE}_{n_2,p_2,q_2}(\bm x) e^{i \omega_2 t}\right], \\
\bm H^{pump}(\bm x, t) &= F_0 \real\left[\bm{\mathcal{M}}^{TM}_{n_1,p_1,q_1}(\bm x) e^{i \omega_1 t} + \bm{\mathcal{M}}^{TE}_{n_2,p_2,q_2}(\bm x) e^{i \omega_2 t}\right].
\end{aligned}
\label{eq:pumps-rect-2a}
\end{equation}
Here the subscript $1(2)$ refers to the TM (TE) mode, $\omega_1 = \omega_{n_1,p_1,q_1}$ and $\omega_2 = \omega_{n_2,p_2,q_2}$. Now the eqs.~\eqref{eq:rhs-rect-1} are to be solved, where the r.h.s.~is computed at the pump field \eqref{eq:pumps-rect-2a}.

Since the r.h.s.~of the wave equation for the signal modes is cubic relatively to the pump modes (see \eqref{eq:corrections}) and the latter are simple trigonometric functions, the r.h.s of \eqref{eq:rhs-rect-1} may contain terms only of the following form:
\begin{equation*}
A\,h(\omega_{sig} t)\,h(k_{sig,x} x)\,h(k_{sig,y} y)\,h(k_{sig,z} z),
\end{equation*}
where the notation $h(\cdot)$ stands just for trigonometrical functions $\sin(\cdot)$ and $\cos(\cdot)$, whereas the signal mode frequency $\omega_{sig}$ and wavevector components $k_{sig,x},~ k_{sig,y},~ k_{sig,z}$ can take arbitrary values at most from the following sets:
\begin{equation}
\label{eq:combs}
\begin{matrix}
\hspace{2mm}\omega_{sig} ~\in~ \{ & \omega_1, \quad & \omega_2, \quad & 2\omega_1 \pm \omega_2, \quad & 2\omega_2 \pm \omega_1, \quad & 3\omega_1, \quad & 3\omega_2 \quad \hspace{1mm}\}, \\
k_{sig,x} ~\in~ \{ & k_{1x}, \quad & k_{2x}, \quad & 2k_{1x} \pm k_{2x}, \quad & 2k_{2x} \pm k_{1x}, \quad & 3k_{1x}, \quad & 3k_{2x} \quad\}, \\
k_{sig,y} ~\in~ \{ & k_{1y}, \quad & k_{2y}, \quad & 2k_{1y} \pm k_{2y}, \quad & 2k_{2y} \pm k_{1y}, \quad & 3k_{1y}, \quad & 3k_{2y} \quad\}, \\
k_{sig,z} ~\in~ \{ & k_{1z}, \quad & k_{2z}, \quad & 2k_{1z} \pm k_{2z}, \quad & 2k_{2z} \pm k_{1z}, \quad & 3k_{1z}, \quad & 3k_{2z} \quad\}. \\
\end{matrix}
\end{equation}
For instance, some mixed combinations like $\sin(3\omega_1)\cos[(2k_{1x}-k_{2x})x]\cos(k_{1y}y)\sin(3k_{2z}z)$ might hypothetically appear within the r.h.s.~of the wave equations \eqref{eq:rhs-rect-1}. However, to search reliably for the resonant components we have to check the conditions of the criterion. The first condition reads 
\begin{equation}
    \omega_{sig}^2 = k_{sig,x}^2 + k_{sig,y}^2 + k_{sig,z}^2.
    \label{square}
\end{equation}

Before the direct test of the second condition, in order to simplify computer algebra computations, we make some additional theoretical statements concerning possible generation of a signal mode with the frequency $\omega_{sig} = 2\omega_1 + \omega_2$. 
First, one can write the triangle inequality for the wavevectors of the pump modes,
\begin{equation}
 \omega_{sig} =   2\omega_1 + \omega_2 =  \left(|2\bm k_1| + |\bm k_2| \right) ~\geqslant~ |2\bm k_1 + \bm k_2| = 
 \sqrt{(2k_{1x}+k_{2x})^2+(2k_{1y}+k_{2y})^2+(2k_{1z}+k_{2z})^2}.
   \label{triangle}
\end{equation}
The equality holds if the two pump mode wavevectors are parallel, $\bm k_1 \parallel \bm k_2$. The condition \eqref{square} for this case is satisfied automatically. In the case of non-parallel wavevectors, the triangle inequality implies that at least one of the components of $\bm k_{sig}$ (say, $k_{sig,x}$) should be larger than $2k_{1x}+k_{2x}$. The only case is the triple maximal projection of wavenumbers, $k_{sig,x} = 3\times \max \left(k_{1x},~k_{2x}\right)$.

Now we are to check the second condition of the resonance criterion. This involves our traditional calculation of $(\bm F, \bm{\mathcal E}_{npq}) = \int_D F_x {\mathcal E}_{npq,x} dV + \int_D F_y {\mathcal E}_{npq,y} dV + \int_D F_z {\mathcal E}_{npq,z} dV$ where $\bm F \in \{\bm F^{el}, ~\bm F^{mg}\}$ and $(npq)$ takes at least $6^3 = 216$ possible combinations from \eqref{eq:combs}. Full symbolic evaluation in \textsc{wxMAXIMA} requires too much computational resources, so we have to truncate out algorithm slightly (only for the current subsection).

Instead of solving full vector system \eqref{eq:problem}, we resort to its component with scalar wave equation for the $z$-component $E^{sig}_z(\bm x, t)$ only. This is generally allowed, since the system \eqref{eq:problem} is already linear with respect to signal fields. Consequently, we write an expansion \eqref{eq:expansion} only for $z$-component:
\begin{equation*}
    E^{sig}_z(\bm x, t) = \sum_{npq} E^{sig}_{z,k} (t) \, {\mathcal E}_{npq,z} (\bm x).
    \label{eq:expansion-scalar}
\end{equation*}

Finally, we obtain second-order differential equation for $E^{sig}_{z,k}(t)$ which is completely similar to \eqref{eq:ode} with the only difference:  its r.h.s.~contains solely the $z$-component of the scalar product $F_k(t) = \int_D F_z(\bm x, t) {\mathcal E}_{npq,z}(\bm x) dV$; the related calculations become $\sim 3$ times simpler than before. Nevertheless, this partial treatment is still sufficient to prove the \emph{absence} of resonant amplification for a certain signal mode: the $z$-component of signal mode induced by arbitrary pump modes should not be zero if such a signal mode were resonantly amplified indeed.

The result of the computer algebra calculations for the temporal spectra of non-zero projections is presented in the simplified form in the Table \ref{tbl:rect-2a}. Note that these spectra relate only to the testing of condition 2, the condition 1 remains still to be examined. Here the signs ``$\pm$'' are independent from each other, so that the table is compressed due to the lack of space. Table \ref{tbl:rect-2a} presents ``an upper limit'' on possible spectra, in specific cases the coefficients before some harmonics vanish to zero. Particularly, these situations include the case of pump modes with parallel wavevectors, $\bm k_1 \parallel \bm k_2$. Thus, it turns out that the corresponding signal mode is not resonantly amplified.
\begin{table}[ht!]
\centering
\def\arraystretch{1.3}
\begin{tabular}{|*{9}{c|}}
    \hline
    Modes: & $\begin{array}{c} n_1 \\ p_1 \\ q_1 \end{array}$ & $\begin{array}{c} 3n_1 \\ p_1 \\ q_1 \end{array}$ & $\begin{array}{c} n_1 \\ 3p_1 \\ q_1 \end{array}$ & $\begin{array}{c} n_1 \\ p_1 \\ 3q_1 \end{array}$ & $\begin{array}{c} n_1 \\ 3p_1 \\ 3q_1 \end{array}$ & $\begin{array}{c} 3n_1 \\ p_1 \\ 3q_1 \end{array}$ & $\begin{array}{c} 3n_1 \\ 3p_1 \\ q_1 \end{array}$ & $\begin{array}{c} 3n_1 \\ 3p_1 \\ 3q_1 \end{array}$ \\
    \hline
    $F^{el}_z~$ on TM & \multirow{2}{*}{$\omega_1, ~ 3\omega_1, ~ 2\omega_2+\omega_1, ~ 2\omega_2-\omega_1$} & \multicolumn{6}{c|}{$\omega_1, ~ 3\omega_1$} & $\omega_1$ \\
    \cline{1-1} \cline{3-9}
    $F^{mg}_z$ on TE & & \multicolumn{2}{c|}{$\omega_1, ~ 3\omega_1$} & \multicolumn{3}{c|}{$\omega_1$} & --- & --- \\
    \hline
\end{tabular} \\
\begin{tabular}{c} \\ \end{tabular} \\
\begin{tabular}{|*{8}{c|}}
    \hline
    Modes: & $\begin{array}{c} 2n_2 \pm n_1 \\ p_1 \\ q_1 \end{array}$ & $\begin{array}{c} n_1 \\ 2p_2 \pm p_1 \\ q_1 \end{array}$ & $\begin{array}{c} n_1 \\ p_1 \\ 2q_2 \pm q_1 \end{array}$ & $\begin{array}{c} 2n_2 \pm n_1 \\ 2p_2 \pm p_1 \\ q_1 \end{array}$ & $\begin{array}{c} 2n_2 \pm n_1 \\ p_1 \\ 2q_2 \pm q_1 \end{array}$ & $\begin{array}{c} n_1 \\ 2p_2 \pm p_1 \\ 2q_2 \pm q_1 \end{array}$ & $\begin{array}{c} 2n_2 \pm n_1 \\ 2p_2 \pm p_1 \\ 2q_2 \pm q_1 \end{array}$  \\
    \hline
    $F^{el}_z~$ on TM & \multicolumn{7}{c|}{\multirow{2}{*}{$\omega_1, ~ 2\omega_2+\omega_1, ~ 2\omega_2-\omega_1$}} \\
    \cline{1-1}
    $F^{mg}_z$ on TE & \multicolumn{7}{c|}{} \\
    \hline
\end{tabular} \\
\caption{Examination of the resonance criterion for two arbitrary pump modes in rectangular cavity.}
\label{tbl:rect-2a}
\end{table}

It addition, it is shown in the Table \ref{tbl:rect-2a} that the sector with mixed wavenumbers and frequencies (the second table among Tables \ref{tbl:rect-2a}) does not interlace with the sector with pump mode or triple wavenumbers and frequencies (the first table among Tables \ref{tbl:rect-2a}). Therefore, the only remaining case of nonparallel wavevectors which require at least one triple wavenumber (see the paragraph after eq.~\eqref{triangle}) is ruled out by the condition 2.

As a result, we have shown by scanning all possible combinations that the signal mode with the frequency $2\omega_1 + \omega_2$ is not resonantly amplified. 

However, as we demonstrate in the following subsection, resonant enhancement does occur for the signal mode of frequency $2\omega_1 - \omega_2$.

\subsubsection{Resonant solution for the $2\omega_1 - \omega_2$ signal mode}

Now we revert to our initial algorithm (not the truncated scalar variant from the previous subsection) and consider two certain pump modes $\rm TM 110 + \rm TE 011$ in order to demonstrate explicitly the resonant amplification of a signal mode with mixed frequency $2\omega_2 - \omega_1$ (where $\omega_1 \equiv \omega_{110}$ and $\omega_2 \equiv \omega_{011}$).

The lowest and the pure triple frequencies trivially match cavity eigenvalues. However, some of $2\omega_{011} \pm \omega_{110}$ and $2\omega_{110} \pm \omega_{011}$ might coincide with intermediate eigenfrequencies too (e.g. with $\omega_{130}$) if one carefully adjusts cavity sides' lengths. Thus, we first project the r.h.s.~of \eqref{eq:rhs-rect-1} on every possible eigenmode from the lowest thru the triple ones (even higher frequencies are guaranteed to be absent in spectra of \eqref{eq:rhs-rect-1}). Temporal spectra of all non-vanishing projections are presented in the Table \ref{tbl:rect-2}.
\begin{table}[ht!]
\centering
\def\arraystretch{1.3}
\begin{tabular}{|l|*{8}{c|}}
    \hline
    TM$(\cdots)$ & $110$ & $130$ & $310$ & $330$ & $112$ & $132$ & $211$ & $231$ \\
    \hline
    $\bm F^{el}$ & \multicolumn{2}{c|}{\multirow{2}{*}{$\omega_{110}, ~ 3\omega_{110}, ~ 2\omega_{011}\pm\omega_{110}$}} & \multirow{2}{*}{$\omega_{110}, ~ 3\omega_{110}$} & \multirow{2}{*}{$\omega_{110}$} & \multicolumn{2}{c|}{\multirow{2}{*}{$\omega_{110}, ~ 2\omega_{011}\pm\omega_{110}$}} & \multicolumn{2}{c|}{\multirow{2}{*}{$\omega_{011}, ~ 2\omega_{110}\pm\omega_{011}$}} \\
    \cline{1-1}
    $\bm F^{mg}$ & \multicolumn{2}{c|}{} & & & \multicolumn{2}{c|}{} & \multicolumn{2}{c|}{} \\
    \hline \hline \hline
    TE$(\cdots)$ & $011$ & $031$ & $013$ & $033$ & $112$ & $132$ & $211$ & $231$ \\
    \hline
    $\bm F^{el}$ & \multicolumn{2}{c|}{\multirow{2}{*}{$\omega_{011}, ~ 3\omega_{011}, ~ 2\omega_{110}\pm\omega_{011}$}} & \multirow{2}{*}{$\omega_{011}, ~ 3\omega_{011}$} & \multirow{2}{*}{$\omega_{011}$} & \multicolumn{2}{c|}{\multirow{2}{*}{$\omega_{110}, ~ 2\omega_{011}\pm\omega_{110}$}} & \multicolumn{2}{c|}{\multirow{2}{*}{$\omega_{011}, ~ 2\omega_{110}\pm\omega_{011}$}} \\
    \cline{1-1}
    $\bm F^{mg}$ & \multicolumn{2}{c|}{} & & & \multicolumn{2}{c|}{} & \multicolumn{2}{c|}{} \\
    \hline
\end{tabular}
\caption{Examination of the resonance criterion for two pump modes in rectangular cavity.}
\label{tbl:rect-2}
\end{table}

As usual, only the lowest harmonics $\omega_{011}$ and $\omega_{110}$ resonate unconditionally. Following the results of the previous subsection, the third harmonics as well as the ``plus'' combined modes do not resonate. The two remaining options (up to the permutation of indices) are $2\omega_{011}-\omega_{110} = \omega_{130}$ and $2\omega_{011}-\omega_{110} = \omega_{132}$. Substituting the expression via wavenumbers (see eq.~\eqref{square}) to the frequency matching conditions, one obtains the conditions to the cavity dimensions $a,b,c$. From one hand, no choice of the cavity dimensions satisfy the second condition $2\omega_{011}-\omega_{110} = \omega_{132}$. From the other hand, the first condition can be satisfied,
\begin{equation}
   2\omega_{011} - \omega_{110} = \omega_{130} \quad \Leftrightarrow \quad \left(\frac{c}{a}\right)^2 \left(\frac{c}{b}\right)^2 + \left(\frac{c}{a}\right)^2 +3\left(\frac{c}{b}\right)^2 = 1.  
   \label{Geom}
\end{equation}
Assuming for simplicity the square section of the cavity $a=b$, we obtain from the eq.~\eqref{Geom} the condition for the cavity length $c$, 
\begin{equation}
    \frac{c}{a} = \frac{c}{b} = \xi = \sqrt{\sqrt{5} - 2} \approx 0.486.
    \label{ResGeom}
\end{equation}

The resonantly enhanced signal mode $\omega_{130}$ (for shortness, $z$-component of the electric field) in the cavity satisfying eq.~\eqref{ResGeom} reads, 
\begin{equation}
     E^{sig}_z (\bm x,t) = B\frac{\pi^2 \epsilon F_0^3 Q }{\omega_{130}^2 c^2 } \sin(\omega_{130}t)\sin\left(\frac{\pi x}{a}\right)\sin\left(\frac{3\pi y}{b}\right),
\end{equation}
where $Q=\omega_{130}/\Gamma$ is the cavity quality factor related to the mode $\omega_{130}$, and the numerical factor $B$ reads,
\begin{equation}
     B = \frac{1}{\xi^2}\biggl(\xi\sqrt{2(1+\xi^2)}+4+\xi^2\biggr) - \frac{\beta}{\xi^2}\biggl(\xi\sqrt{2(1+\xi^2)^3}+1-\xi^2\biggr) \ \approx \ 8.517 \ \ \ \mbox{for}\ \beta=\frac{7}{4}.
     \label{Coeff}
\end{equation}

Note that for a certain critical $\beta \approx 2.92$ the signal mode vanishes even for the resonant cavity geometry. However, for other choice of pump and signal modes (say, $\omega_{sig}=\omega_{150}=2\omega_{130}-\omega_{011}$) the resonant geometry configuration would be changed, and the numerical value of critical $\beta$ would be different.

It seems to be a counter-intuitive result that one can resonantly enhance signal mode $2\omega_2 - \omega_1$ (after certain adjustment of cavity geometry), whereas the mode $2\omega_2 + \omega_1$ remains always suppressed.

\section{Discussion}
\label{sec:discuss}

In the current paper we have formulated the conditions for the resonant amplification of a signal mode in a cavity of arbitrary shape, and applied them to the analysis of linear and rectangular cavities. We have demonstrated that two pump modes with frequencies $\omega_1$ and $\omega_2$ in a rectangular cavity resonantly produce a signal mode with frequency $2\omega_2 - \omega_1$ ($2\omega_1 - \omega_2$) for a certain cavity geometry. On the other side, we have proved that the signal modes with frequencies $2\omega_1 + \omega_2$ ($2\omega_2 + \omega_1$), as well as the third harmonics $3\omega_{1}$ ($3\omega_{2}$) are not resonantly amplified. Remind that the resonant amplification means an enhancement in $Q$ times, where $Q$ is the cavity quality factor which can achieve a numerical value up to $10^{12}$ \cite{Romanenko:2018nut}.

The crucial point of our proof for the absence of resonance in rectangular cavity is that the cavity eigenmodes include only trigonometric functions which allows us to make analytic calculations for arbitrary cavity modes. This stops working for a cavity of arbitrary shape, in that case the numerical calculations for specific cavity modes become necessary.

The absence of resonant amplification of the third harmonics and combined ``plus'' modes seems to be connected with the polarization structure of the vector gauge field. Considering the similar problem of higher-order harmonics generation for massless scalar field with $\lambda \phi^4$ interaction instead of the electromagnetic one, one first obtains the analogue of the inhomogeneous wave equation (\ref{eq:mod-waves}) which reads $\Box \phi = \lambda \phi^3$. Decomposing $\phi$ into the pump mode $\phi_{pump}$ which is a cavity eigenmode, and the signal mode $\phi_{sig}$ of small amplitude, one can easily check that the third harmonics does generate for the scalar field. We hope that the reason of this difference between scalar and electromagnetic field would become more clear after considering aforementioned processes on quantum level.

To conclude, we have considered the problem of vacuum generation of higher-order harmonics only from the theoretical side. Although the scheme of such experiment for the case of cylindrical cavity was in general studied in \cite{Bogorad:2019pbu}, there are still several unsolved issues, including proper treatment of nonlinearities from the cavity walls etc.

\paragraph{Acknowledgments} The Authors thank Maxim Fitkevich, Dmitry Kirpichnikov, Dmitry Levkov, Valery Rubakov and Dmitry Salnikov for helpful discussions. The work is supported by RSF grant 21-72-10151.

\bibliographystyle{unsrt}
\bibliography{biblio}

\end{document}